\documentclass[onecolumn]{revtex4}

\input{epsf}

\def\AJ{{\it Ap. J.} }    
 
    \def\CQG{{\it Class. Quantum Gravity} }
    \def\IJMP{{\it Int. J. Mod. Phys.} }

  \def\PR{{\it
Phys. Rev.} } \def\PRL{{\it Phys. Rev. Lett.} }

\def\al{\alpha}   
   
\def\th{\theta}   
\def\la{\lambda}

  \def\De{\Delta}

 \def\frac#1#2{{\textstyle{{#1}\over
{#2}}}} 
\def\lsim{\mathrel{\rlap{\lower4pt\hbox{\hskip1pt$\sim$}}
\raise1pt\hbox{$<$}}}
\def\gsim{\mathrel{\rlap{\lower4pt\hbox{\hskip1pt$\sim$}}
\raise1pt\hbox{$>$}}} \def\sqr#1#2{{\vcenter{\vbox{\hrule height.#2pt
\hbox{\vrule width.#2pt height#1pt \kern#1pt \vrule width.#2pt} \hrule
height.#2pt}}}}

\def\beq{\begin{equation}} \def\eeq{\end{equation}}
\def\beqa{\begin{eqnarray}} \def\eeqa{\end{eqnarray}}

\begin{document}

\title{A mission to test the Pioneer anomaly: estimating the main
systematic effects}

\vskip 0.2cm

\author{O. Bertolami \footnotemark, J. P\'aramos \footnote[2]{Also at
Centro de F\'{i}sica dos Plasmas, Instituto Superior T\'ecnico,
Lisbon.}}

\vskip 0.2cm

\affiliation{Instituto Superior T\'ecnico, Departamento de
F\'{\i}sica, \\ Av. Rovisco Pais 1, 1049-001 Lisboa, Portugal}

\vskip 0.2cm

\affiliation{E-mail addresses: orfeu@cosmos.ist.utl.pt,
x\_jorge@fisica.ist.utl.pt}

\vskip 0.5cm

\date{\today}

\begin{abstract}

We estimate the main systematic effects relevant in a mission to test
and characterize the Pioneer anomaly through the flight formation
concept, by launching probing spheres from a mother spacecraft and
tracking their motion via laser ranging.

\vskip 0.5cm

\end{abstract}

\pacs{ \hspace{2cm}Preprint DF/IST-x.2007}

\maketitle


\section{Introduction} Analyses of radiometric data of the Pioneer 10
and 11 spacecraft do suggest the existence of an anomalous
acceleration on the two spacecraft, inbound to the Sun and with a
(constant) magnitude of $a_P \simeq (8.5 \pm 1.3) \times 10^{-10} ~
m/s^2 $.  Extensive attempts carried out so far to explain this
phenomena as a result of an estimate of the main systematic effects
accounting for thermal and mechanical effects, as well as errors in
the tracking algorithms used, have shown to be unsuccessful
\cite{JPL}, despite claims otherwise \cite{Lou}. Also, another
analysis has shown that a secular trend in the anomaly may be found,
with a time constant larger than 50 years allowed: this still leaves
room for thermal radiation to account for the Pioneer anomaly (given
the $\sim88$ years half-life of the plutonium source in the
radio-thermal generators, which should be somewhat lowered due to
degradation of the thermal coupling), and is undergoing further
studies by groups within the Pioneer collaboration team
\cite{team}. The two Pioneer spacecraft follow approximate opposite
hyperbolic trajectories away from the Solar System but, despite that,
the same anomaly was found. This prompts for an intriguing question:
what is the fundamental, and possibly new, physics behind this
anomaly?

To answer this, many proposals have been advanced, as summarized in
Ref. \cite{Paramos}. However, before the possibility for new physics
is seriously addressed, an unequivocal description of the anomaly, as
given by already available data, should be given. In particular,
secular and spatial trends should be carefully modeled, both from a
statistical point of view and as possible thermal and engineering
causes of the anomalous acceleration. Furthermore, the direction of
the acceleration vector must be clearly characterized. Indeed, the
distances involved in the already conducted Doppler analysis do not
allow for an angular discrimination between the several candidates,
each remarking a different origin for the anomaly: a line of action
pointing towards the Sun would indicate an effect of gravitational
origin (since solar pressure is manifestly too low to account for the
effect), while a Earth-bound anomaly would indicate either a modified
Doppler effect (and hence, new physics affecting light propagation and
causing an effective blue shift), or an incorrect modeling of Doppler
data, possibly due to Deep Space Network and software clock drifts,
incorrect ephemerides estimates, mismodeled Earth orientation
parameters, {\it etc}; the anomaly may also point along the spin axis
of the spacecraft, indicating that onboard, underestimated systematic
effects are responsible for it; an anomaly along the velocity vector
would hint at some sort of drag effect, possibly related to dark
matter or dust distribution (although these are currently known to a
good accuracy, and yield much lower effects), or to a modification of
geodetical motion also hinting at a modification or extension of
General Relativity.

Clearly, it is difficult to correctly assign a definitive origin to
the observables of interest: the direction, magnitude, spatial and
secular variation of the anomaly. This further enforces the need to
carefully account for all known effects, despite the added difficulty
of dealing with data that is almost thirty years old! For sure, rather
complex and subtle model-dependent computations must be carried out -
preferably starting with different assumptions and hopefully getting
the same final answer for the enigma of the Pioneer anomaly.

\section{Acceleration budget}

In order to test the Pioneer anomaly, one must fully characterize its
features by removing the uncertainties due to systematics and other
biases, and ascertain the magnitude and direction of the anomalous
acceleration. Thus, we begin by listing the various components that
may affect an extended object, a spacecraft, traveling through the
solar system:

\beq {\bf F} = {\bf F}_{Grav} + {\bf F}_{Pres} + {\bf F}_{Drag} + {\bf
F}_{L} + {\bf F}_{Syst} + {\bf F}_{Unk} ~~, \label{comp} \eeq

\noindent where the underscripts indicate the origin of the various
components. The gravitational force ${\bf F}_{Grav}$ is given by the
sum of the Newtonian forces corresponding to all the major bodies in
the solar system, yielding about $a_{Grav} = \Sigma _i GM_im/r_i^2= 6
\times 10^{-3} ~m/s^2$ at a distance from the Sun of $1~AU$. The solar
pressure ${\bf F}_{Pres}$ is divided into a radiation component, given
by $K f_\odot A cos \th(r) / c r^2 $, where $c$ is the speed of light,
$K$ is the effective absorption/reflection coefficient (of order
unity), $f_\odot$ is the effective-temperature ``solar radiation
constant'' -- approximately $1.4~kW/m^2 (AU)^2$ at $ r = 1 ~AU$ --,
$A$ is the projected area to the solar vector, $\th$ is the angle
between the spacecraft antenna's axis and the direction of the Sun,
and $r$ is the distance from the Sun; for a cross-sectional area of
the order of $10~m^2$, this yields approximately
$a_{Pres}=10^{-8}-10^{-7}~m/s^2$ (depending on assumptions) at
$r=1~AU$, and falls below the reported magnitude of the Pioneer
anomaly at $r \approx 11 ~AU$ (for the upper bound, or a much closer
$r \approx 3.5~AU$ for the lower bound) -- well within the flight
envelope of a typical deep space mission. The solar wind component is
given by a similar expression, with $f_\odot$ substituted by $m_p v^3
n$, where $ n \approx 5~cm^{-3}$ is the proton density at $1~AU$ and
$v \approx 400 ~km/s$ is the solar wind speed, this yields a value of
the order $a_{wind}=10^{-13}-10^{-12}~m/s^2$.  The drag force due to
interplanetary dust is given by $- k A \rho(r) v(r)^2 \hat{\bf v}$,
where $k$ is a characteristic constant of order unity, $A$ is the
cross-section, $\rho \approx 10^{-24}~g/cm^3$ is the interplanetary
medium density and $v \approx 400 ~km/s$ is the body's velocity; this
yield approximately $a_{Drag} = 10^{-12}-10^{-11}~m/s^2$. The
electromagnetic Lorentz force, related to local magnetic field ${\bf
B}$, the probe's charge $q$ and velocity ${\bf v}$ by $ {\bf F}_{L} =
q {\bf v} \times {\bf B}$; although the accumulated charge in the
spacecraft and the magnetic field may vary widely, a upper bound of
$a_{L} \sim 10^{-12}~m/s^2$ is a typical value \cite{Null}. Finally,
the systematic component accounts for possible gas leakage,
anisotropic heat emission and reflectance, antenna recoil and other
effects; and, finally, we consider a component of unknown origin.

The analysis of the relevant systematic components can be simplified
by comparing the precision of acceleration measurements with the order
of magnitude of the above components: when the former is higher than
the latter, one can safely disregard them from the overall picture. In
the following, we assume that the low systematics intended for a
dedicated probe allow for a Doppler tracking accuracy of order
$10^{-11}~m/s^2$, and a direct measurement through internal
accelerometer of order $10^{-12}~m/s^2$. The former is one order of
magnitude lower than the accuracy available for the Pioneer data, and
the latter corresponds to state of the art low-frequency
accelerometers.

An obvious statement concerning any dedicated mission to test the
Pioneer anomaly is that it should be more robust than these scientific
probes, that is, that the preceding components for the Doppler and
acceleration measurements should be of smaller magnitude than those
affecting the Pioneer spacecraft. Hence, from Table II of
Ref. \cite{JPL}, we get the following upper limits for these
components:


\begin{table*}[ht] \begin{center} \caption{Error budget for the
Pioneer 10 and 11, taken from Ref. \cite{JPL}.}

\label{table}

\vskip 20pt

\begin{tabular}{rlll} Item & Description of error budget constituents
& Bias~~~~~& Uncertainty \\ & & $10^{-8} ~\rm cm/s^2$ & $10^{-8} ~\rm
cm/s^2$ \\\hline & & & \\ 1 & {\sf Systematics generated external to
the spacecraft:} && \\ & a) Solar radiation pressure and mass &
$+0.03$ & $\pm 0.01$\\ & b) Solar wind && $ \pm < 10^{-5}$ \\ & c)
Solar corona & & $ \pm 0.02$ \\ & d) Electro-magnetic Lorentz forces
&& $\pm < 10^{-4}$ \\ & e) Influence of the Kuiper belt's gravity &&
$\pm 0.03$\\ & f) Influence of the Earth orientation && $\pm 0.001$ \\
& g) Mechanical and phase stability of DSN antennae && $\pm < 0.001$
\\ & h) Phase stability and clocks && $\pm <0.001$ \\ & i) DSN station
location && $\pm < 10^{-5}$ \\ & j) Troposphere and ionosphere && $\pm
< 0.001$ \\[10pt] 2 & {\sf On-board generated systematics:} && \\ & a)
Radio beam reaction force & $+1.10$&$\pm 0.11$ \\ & b) RTG heat
reflected off the craft & $-0.55$&$\pm 0.55$ \\ & c) Differential
emissivity of the RTGs & & $\pm 0.85$ \\ & d) Non-isotropic radiative
cooling of the spacecraft && $\pm 0.48$\\ & e) Expelled Helium
produced within the RTGs &$+0.15$ & $\pm 0.16$ \\ & f) Gas leakage & &
$\pm 0.56$ \\ & g) Variation between spacecraft determinations &
$+0.17$ & $\pm 0.17$ \\[10pt] 3 & {\sf Computational systematics:} &&
\\ & a) Numerical stability of least-squares estimation & &$\pm0.02$
\\ & b) Accuracy of consistency/model tests & &$\pm0.13$ \\ & c)
Mismodeling of maneuvers & & $\pm 0.01$ \\ & d) Mismodeling of the
solar corona & &$\pm 0.02$ \\ & e) Annual/diurnal terms & & $\pm 0.32$
\\[10pt] \hline & && \\ & Estimate of total bias/error & $+0.90$& $\pm
1.33$ \\ & && \\ \end{tabular} \end{center} \end{table*}


As stated in Ref. \cite{JPL}, one can safely assume that the anomalous
acceleration is not due to electromagnetic forces, nor to solar
radiation and solar wind pressures. Also, it is argued that the drag
due to the surrounding environment is not enough to account for the
anomaly; regarding this, one may ask what should be the medium's
density, so that a $v^2$ dependent drag force would account for the
anomaly: a simple calculation shows that this should be of order
$10^{-19}~g/cm^3$; for comparison, the density of interplanetary dust,
arising from hot-wind plasma \cite{Kimura}, is lesser than
$10^{-24}~g/cm^3$; the density of interstellar dust (directly measured
by the Ulysses spacecraft) is even smaller, at about $ 3 \times
10^{-26}~g/cm^3$.

Also, an under-estimated mass of the yet unexplored Kuiper belt could
possibly account for the anomalous acceleration. However, it has been
shown that this would require an abnormally high mass for this
extended object, about two order of magnitude higher than the commonly
accepted value of $M_{Kuiper} = 0.3 M_{Earth}$ \cite{JPL,Vieira,
Nietokuiper}.

The above discussion supports the acknowledged status of the Pioneer
anomaly, as currently stated by the Pioneer anomaly collaboration: the
origin of the anomaly can be due to systematics, such as
underestimated thermal effects, or to new physics. Clearly, the
presented error budget shows that the systematics errors cannot
account for the anomaly; however, this figure was obtained resorting
to the available data concerning RTG power decay and emissivity, radio
antenna emissions, Helium leak in the RTGs, amongst other competing
effects (see Ref. \cite{JPL} for a full discussion). In principle, it
is conceivable that a yet unaccounted systematic effect can explain
the anomaly. For this reason, it is not sufficient to design a
dedicated probe in such a way as to minimize known systematic effects,
namely through a careful thermal modeling and symmetrical geometry;
indeed, a key feature of the proposed dedicated probe \cite{team} is
that it allows for a direct assessment of these systematic
perturbations.

As proposed in Ref. \cite{team}, this is attained thanks to the direct
measurement of the distance between the primary craft and a small
passive sphere -- circa $10~cm$ diameter and with no attitude or
stabilization capability. This could be attained either by endowing
the latter with a low power transponder, or by covering it with
cornercube retroreflectors, allowing for laser-ranging from the
primary craft -- these devices reflect any incoming light ray in the
incoming direction, eliminating the need for a perfect alignment
between the laser sight and the passive sphere. Asides from design
simplicity, a small sphere allows for a better characterization of its
surface and thermal properties, and also yields smaller and more
easily modeled temperature gradients; also, the lack of attitude
control allows the measurement of accelerations that could otherwise
be below any offset imposed by thruster leaks, mismodeled maneuvers,
{\it etc.}. In order to stay clear from the radio beam, and also to
provide shielding from solar radiation and wind, this sphere is
ejected towards the ``front'' of the primary craft; in order to
provide for the smoothest release possible, its drift speed should be
relatively low, say $1~mm/s$ to $1~cm/s$. Furthermore, as in the case
of the Pioneer probes, the primary craft is not three-axis stabilized,
but instead spins around its axis of inertia: this allows for an
averaging of off-axis systematic effects and provides a ``cleaner''
signal than would be available from the latter stabilization method,
which would require a careful modeling of micro-thrust maneuvers and
effectively reduce the measurement sensitivity (this should be the
reason why the Pioneer anomaly has only been detected on
spin-stabilized probes -- the Galileo and Ulysses missions, aside from
the two homonymous explorers).

We remark that this proposal, although ideally implemented as a
dedicated mission, could also be deployed as a ``piggy-back'' payload
onboard another deep space probe; this would help to circumvent
funding constraints, reduce overall costs and strengthen the
scientific scope of both missions. Clearly, the trade-off is a
decrease in tracking precision of the primary craft, since it would
encompass other scientific instruments and more complex design,
increasing the noise level of both the Doppler tracking and the
onboard acceleration measurements.

We now ascertain the magnitude of the various components present in
Eq. (\ref{comp}), regarding this passive sphere. We do this by scaling
the upper bounds presented in Table \ref{table} (a more rigorous, but
less intuitive description of the various components can be found
elsewhere \cite{Chui}; all results are mutually compatible). We assume
a radius of $10~cm$ and a mass of $1~kg$ for the sphere, and a radius
of $1~m$ and mass of $200~kg$ for the primary craft. With these
quantities, we may scale the relevant acceleration components and
obtain an order of magnitude budget of the systematic effects
affecting the passive sphere:

i) Solar radiation and pressure: we assume the thermal emissivity of
the sphere to be of the same order of magnitude or lower than that of
the primary craft; given the area dependence, we obtain

\beq a_{Pres. sphere} = \left({r_{sphere} \over r_{craft}}\right)^2
{m_{craft} \over m_{sphere}} a_{Pres. craft} = 2 a_{Pres. craft} < 6
\times 10^{-12}~m/s^2 ~~. \eeq

\noindent Hence, this component cannot be mistaken as an anomalous
acceleration with magnitude $a_P = 8 \times
10^{-10}~m/s^2$. Furthermore, one could exploit the advantageous cover
(but not essential, given the low value of $a_{Pres. sphere}$) that
the primary craft provides: the sphere is effectively shielded from
these acceleration components by the ``shadow cone'' of the primary
craft, up to a distance $d$ given by the inequality $d / r_{craft} = r
/ R_\odot $, where $r$ is the distance from the primary craft to the
Sun and $R_\odot$ the radius of the Sun, from which we get $ d \approx
100 (r / 1 ~AU)$. Assuming a minimum clear distance (that is,
sufficiently far away from the primary craft) of $50~m$, we conclude
that the solar radiation and pressure are blocked by the primary craft
after it has achieved a distance from the Sun of $5~AU$, close to
Jupiter's orbit. Recall that the solar pressure affecting the primary
craft falls below the reported magnitude of the Pioneer anomaly at $r
\approx 4 - 10~AU$: a mission profile would possibly encompass a
low-power initial phase before this distance is achieved, with
calibration and other tests performed -- followed by a second phase,
in which high-precision measurements would be conducted.

ii) Drag force: this force is proportional to the square of the
velocity of the sphere with respect to the surrounding medium, and to
its cross-section; given that the drift speed from the primary craft
is very low compared to the latter's speed, one obtains the same
scaling law as in the above case and, therefore, one can exclude drag
forces as a competing effect with the anomalous acceleration under
scrutiny.

iii) Electromagnetic forces: here we address the Lorentz forces due to
interaction with cosmic rays, planetary and solar magnetic fields and
charged plasma in the vicinity of the probe; these are proportional to
the charge of the sphere, and also to its speed, which is approximated
by that of the primary craft. Considering any electric charge to be
evenly distributed throughout the volume of the probe, prior to the
ejection of the sphere, we obtain the scaling law

\beq a_{EM sphere} = (r_{sphere} / r_{craft})^3 (m_{craft} /
m_{sphere}) a_{Lorentz craft} = 0.2 a_{Lorentz craft} \leq
10^{-15}~m/s^2~~;\eeq

\noindent this indicates that electromagnetic forces affecting the
sphere may be neglected.

iv) Electrostatic force due to primary: assuming that the probe is
charged with a total charge $Q$, evenly distributed, one may estimate
the charge of the passive sphere as $Q_{sphere} \approx (r_{sphere} /
r_{craft})^3 Q_{craft} \approx (r_{sphere} / r_{craft})^3 Q $. Hence,
the electrostatic force between the passive sphere and the primary
craft, at a distance d, is given by

\beq F_{Coulomb} = {1 \over 4 \pi \epsilon_0} \left( { r_{sphere}
\over r_{craft} } \right)^3 \left( { Q \over d } \right)^2 ~~. \eeq

\noindent Charge measurements on spacecraft are usually referred by
the equivalent induced voltage $U$; if we simplistically assume a
spherical body of radius $r$, one gets $ U = Q / 4 \pi \epsilon_0
r$. Hence, the above simplifies to

\beq a_{Coulomb} = {F_{Coulomb} \over m_{sphere}} = 4 \pi \epsilon_0
{r_{sphere}^3 \over m_{sphere} r_{craft} d^2} U^2 \approx 1.1 \times
10^{-13} \left({U \over 1~V }\right)^2 \left( { d \over
1~m}\right)^{-2}~m/s^2~~. \eeq

\noindent Considering a minimum distance of $50~m$ shows that the
acceleration of the passive sphere due to the Coulomb force is smaller
than $10^{-12}~m.s^{-2}$ for a primary craft's potential $U$ below
$23~kVolt$, and smaller than the reported magnitude of the Pioneer
anomaly below a potential as large as $19~MVolt$. Notice that the
primary craft is also attracted towards the passive sphere; however,
since this acceleration is much smaller than the inverse (by a factor
$m_{sphere}/m_{craft} = 200$), the relative acceleration is well
approximated by the above. The first obtained limit, although
extremely high, may sometimes occur in space environs, particularly
during nightside magnetospheric storms \cite{charging}, where the
potential can reach approximately $20~kVolt$.

From the above, some guidelines for controlling the effect of Coulomb
forces may be sketched: firstly, to guarantee that the ejection of the
passive sphere occurs at a mild magnetic environment, in the absence
of increased solar activity and away from magnetic belts of planets
and other sources. Also, the probe should encompass a device to
measure its own electric potential. Furthermore, instead of an overall
control of the probe's charge, one may simply control the charge
deposited in the passive sphere. Indeed, the above reasoning assumed
an evenly distributed (volume) charge; even without charge balancing
of the primary craft, any reduction of the sphere's charge would
amount to an equal reduction in the magnitude of the related
acceleration.

Although one lacks the proper knowledge to fully address this issue,
two mechanisms may be outlined to diminish the charge deposition in
the passive sphere; firstly, field-emission cathodes may be used to
electrically neutralize it, in a similar fashion to the method used to
control the gyroscopes of the Gravity Probe B \cite{GPB}. Secondly,
the passive sphere may be discharged by contact, prior to release, by
establishing a parallel circuit with a capacitor with a much larger
capacity; assuming that the passive sphere has a dielectric constant
of approximately $\epsilon_r = 3.75$ (value for a corner cube with
fused silica), its capacity is $ C= 4 \pi \epsilon_0 \epsilon_r
r_{sphere} \approx 20~pF$; a capacitor of just $1~nF$ would diminish
the sphere's charge by a factor of $98\%$.

v) Thermal radiation from the primary craft: given that the radio beam
points oppositely to the ejected sphere, one should only account for
thermal radiation arising from the reflected emission of the RTGs and
cooling of the primary craft; following Ref. \cite{JPL}, we assume
that the dissipated power from the main hub has an upper bound of
order $100~W$; also, we assume that the RTGs emit approximately
tenfold. However, the design of a dedicated probe should minimize
this, by concealing the RTGs into the geometry of the craft; hence, we
assume that only a fraction of $10\%$ (an extremely generous
estimative) is reflected onto the sphere. Assuming isotropy, the
sphere is subjected to an incident power given by $ P =
110~W~(r_{sphere} / d)^2 $; for distances larger than $500~m$, this is
smaller than $1.1 ~\mu W$. The ensuing acceleration is then
$a_{thermal} = P / m_{sphere} c \leq 3.7 \times 10^{-15}~m/s^2$, and
can thus be neglected.

vi) Recoil from the laser ranging: given the above paragraph, a laser
ranging unit with power of the order of $1 mW$ directly focused at the
passive sphere imposes an acceleration of $a_{laser} = P / m_{sphere}
c \leq 3.3 \times 10^{-12}~m/s^2$; this is possibly within the
accuracy of the measurement, but clearly below the necessary magnitude
to compete with the anomalous acceleration.

From the above, we conclude that all environmental and known
systematics may be factored out of the problem, so that one must only
be concerned with unaccounted systematics and forces manifesting new
physics, regarding both the passive sphere and the primary
craft. Hence, the relative acceleration between them is given by

\beq {\bf a}_R \equiv {\bf a}_{craft} - {\bf a}_{sphere} = { {\bf
F}_{Syst. craft} \over m_{craft} } + { {\bf F}_{Unk. craft} \over
m_{craft} } - { {\bf F}_{Unk. sphere} \over m_{sphere} }
\label{relative} ~~,\eeq

\noindent where ${\bf F}_{Syst. craft}$ refers to possible,
unaccounted systematic effects that may account for the Pioneer
anomaly.

If the Pioneer anomaly is a real effect, imposing an equal
acceleration upon bodies of different mass or composition, then the $
{\bf F}_{Unk}$ terms should cancel, and $ {\bf a}_R = {\bf
F}_{Syst. craft} / m_{craft}$: this allows for a direct discrimination
of the systematic effects affecting the primary craft, which could
offer a solution for the Pioneer anomaly without the need of new
physics.

\section{Physically anomalous Doppler tracking}

Asides from a magnitude of $a_P = ( 8.74 \pm 1.33 ) \times
10^{-10}~m/s^2$, the reported anomalous acceleration is widely
credited as an attractive, sunbound effect. However, the determination
of the direction of the anomaly is troublesome, given that the angular
precision of the Pioneer probes was insufficient to distinguish
between the different possibilities: a sunbound physical acceleration,
an Earthbound effect due to misunderstood Doppler tracking or some
other effect, an axis of inertia directed acceleration, reflecting
some overlooked engineering or thermal effect, etc. A thorough
analysis of the earlier stages of the Pioneer missions may clarify
these issues, and the increased precision of a dedicated probe would
certainly work towards that objective.

There is, however, a resilient ambiguity concerning the determination
of the direction of the anomaly. Indeed, if it is found that the
anomalous acceleration points towards the Earth, thus excluding a
potential gravitational origin, one question remains: is it due to
some unaccounted effect due to the power radiated from the aligned
antenna (or other equipment, given that any thermal radiation along
the axis of inertia should be aligned with it? Or is the anomaly due
to a poorly understood Doppler effect? In fact, the latter issue is
more general, since the current sunbound direction is also
extrapolated from Doppler tracking. Here, the question is not if the
trajectory was correctly analyzed and compared with ephemerides
models, but if the actual underlying physical mechanism behind the
tracking, the Doppler effect, is properly interpreted. This is a
commonly neglected concern in the available literature, which tends to
allow only for new physics of gravitational origin, with the Sun as
source; however, an anomalous Doppler effect would still be of great
theoretical interest, hinting further towards a better understanding
of General Relativity and, possibly, future extensions and competing
theories.

Given that the laser ranging used for tracking of the passive sphere
relies on the ``bounce'' time of the emitted pulses, it should not be
affected by any anomalous Doppler shift. Hence, if this is the origin
of the Pioneer anomaly, it should not be probed by the measurement of
the relative acceleration between the passive sphere and the primary
craft:, the $ {\bf F}_{Unk sphere} $ term should vanish, while the
${\bf F}_{Unk craft}$ should subside, instead of canceling each other
in the relative acceleration. In order to strengthen the test of this
hypothesis, we believe that an independent low frequency accelerometer
should be a priority payload of the primary craft. If this is the
case, not only will an anomalous Doppler shift be shown by the
measurement of the passive sphere's relative acceleration, but the
difference between measurements of the acceleration of the primary
craft through Earth-based Doppler tracking and the onboard
accelerometer would also reveal this effect. Notice, however, that an
accelerometer will show no anomalous signal if the Pioneer anomaly is
of gravitational origin, since both the spacecraft and the device's
test mass will be subject to the same acceleration.

\section{Beyond the Pioneer anomaly}

Although the proposed dedicated mission to test the Pioneer anomaly
has been viewed from this specific point of view, it can be more
generally regarded as a probe to measure the acceleration profile of
the Solar System. In this sense, one can speak of a scientific
objective beyond the Pioneer anomaly: the accurate scrutiny of very
low magnitude effects in widely different environs of the solar
system. We now address some of the possibilities that such an
instrument may offer us.

\subsection{A Yukawa force}

As was noted early on \cite{JPL}, a possible interpretation of the
Pioneer anomaly is that it is due to an added Yukawa force, possibly
due to to the dynamics of a massive scalar/vector field with the Sun
as source. The full gravitational potential may then be written as

\beq V(r) = -{G M_\odot \over (1 + \al)r } \left(1 + \al e^{-r/\la}
\right)~~, \eeq

\noindent where $\al$ is the coupling strength and $\la$ is the range
of the Yukawa component, inversely proportional to the second
derivative of the scalar potential affecting the dynamics of the
field, taken at its vacuum expectation value. An expansion of the
derived acceleration $a = -dV / dr$ shows that, asides from the usual
inverse square law Newtonian term, one obtains a constant term which
may be identified with the Pioneer anomalous acceleration, $a_P = -a_N
\al \left[2(1+\al)\right]^{-1} (1~AU/\la)^2$, where $a_N = GM_\odot /
(1~AU)^2 = 5.93 \times 10^{-3}~m.s^{-2} $ is the Newtonian
acceleration at the distance of $1~AU$. This identification yields the
curve $\al(1+\al)^{-1}(1~AU/\la)^2 = - 2 a_P / a_N = - 2.95 \times
10^{-7}$; one remarkable feature is the negative coupling strength,
which is characteristic of massive vector fields.

Typically, one chooses the lengthscale of the Yukawa force $\la$, and
derives its strength from the solution curve, so that $\al \ll
1$. However, little attention has been paid to the remaining terms of
the Taylor expansion; these can be written as $a_n = C_n a_P
(r/\la)^n$, with $n=1,~2,...$ and $C_n < 1$. Assuming $ \la > r$ in
the region of interest, the next to leading order is the term $a_1 =
(2/3) a_P (r/\la)$; for distances between $R_i = 20~AU$ and $R_o =
67~AU$, this linear term must be smaller than the allowed error margin
for the reported anomalous acceleration, which amounts to $ 15.2 \%$
of its center value. Thus, we obtain \beq {2 \over 3} {a_P \over \la}
\leq {2 \times 0.152 a_P \over R_o - R_i} ~~, \eeq

\noindent which implies that $\la \geq 2.19 (R_o-R_i) \simeq 103 ~AU$.

\subsection{A probe of the Kuiper belt mass distribution}


\begin{figure}

\epsfysize=8cm \epsffile{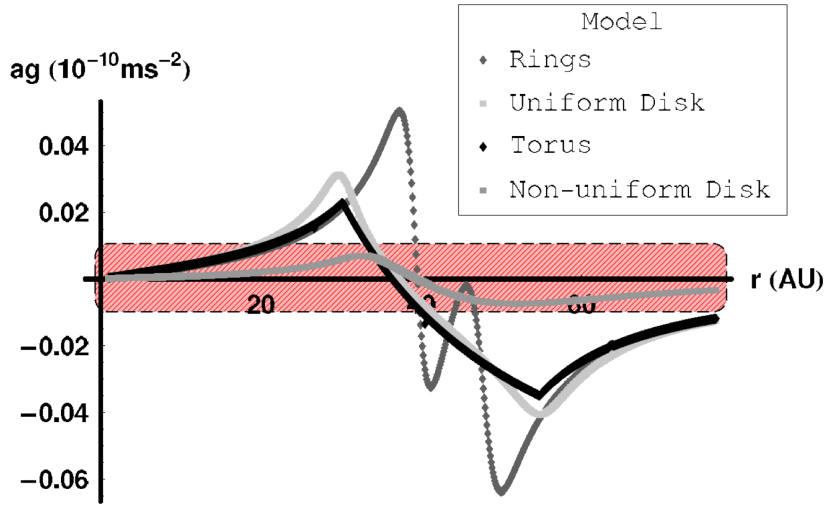} \caption{Kuiper belt mass
distributions, from Ref. \cite{Vieira}, with an accuracy threshold of
$10^{-12}~m/s^2$ superimposed.}  \label{kuiperfig}

\end{figure}



\begin{figure}

\epsfysize=8cm \epsffile{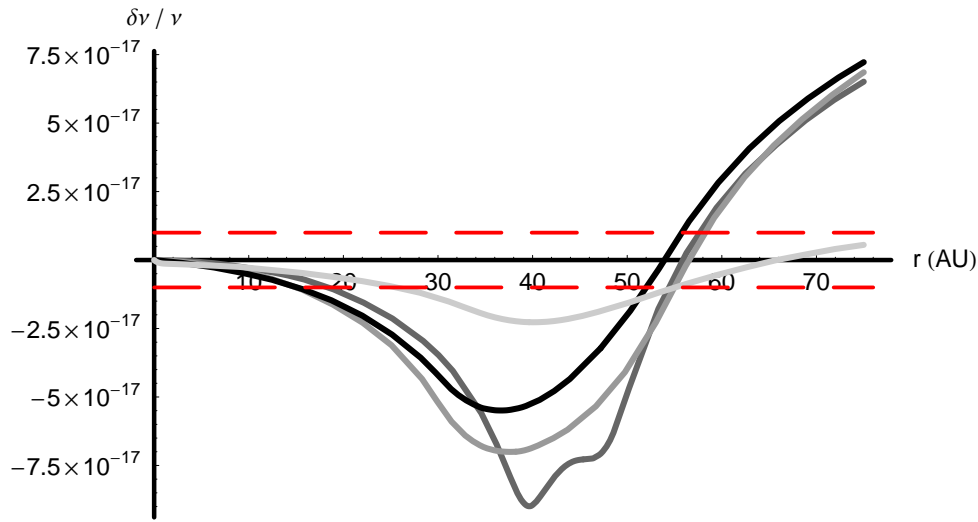} \caption{Relative frequency
shift due to different Kuiper Belt mass distributions for a solar
ecliptic latitude of $\theta=3^{\circ}$, with an accuracy threshold of
$10^{-17} $ superimposed (same labels as in Fig. \ref{kuiperfig}).}
\label{kuiperfig2}

\end{figure}


As shown in Ref. \cite{Vieira,Nietokuiper}, the mass distribution of
the Kuiper belt cannot account for the Pioneer anomaly, providing
neither its magnitude nor its constant behaviour (or allowed spatial
variation \cite{Craig}). However, the acceleration sensitivity of the
discussed dedicated probe could, in principle, be of great use in
discriminating between different models for the spatial distribution
of the estimated $0.3~M_{Earth}$ Kuiper belt. Indeed, by plotting the
different acceleration profiles of competing models (taken from
Ref. \cite{Vieira}), against the assumed $10^{-12}~m/s^2$ accuracy in
Fig. \ref{kuiperfig}, one concludes that the former should be
detectable for a wide range. Specifically, one gets the following
``positive signature'' intervals, where gravity from the Kuiper belt
could be measured with the onboard accelerometer: for the two rings
mass distribution, this starts at $23~ AU$, with gaps at $~ 40~ AU$
and $41 - 47 ~AU$; for the uniform disk mass distribution, this starts
at $ 22 ~ AU$, with a gap at $34 - 43~ AU$; for the toroidal mass
distribution, this starts at $ 22.5~ AU$, with a gap at $33 - 42~ AU$;
finally, the non-uniform disk model is outside the limit of
measurability, although its acceleration peaks at $32$ and $48~ AU$
could perhaps be detected by a careful analysis, particularly if the
total mass of the Kuiper belt is slightly higher than acknowledged.

Alternatively, one could rely on precise Doppler tracking and, by
measuring relative frequency shifts, directly probe the gravitational
potential $U = c^2 \De \nu / \nu$. By plotting the
gravitational potential derived from different models for the Kuiper
belt's mass distribution against an assumed accuracy of $10^{-17}$ in
Fig. \ref{kuiperfig2}, we obtain the following positive signature
intervals: the two rings mass distribution is detectable at the
interval $18~AU < r < 54~AU$ and for $r > 58~AU$; the uniform disk
mass distribution is detectable at the interval $15~AU<r<55~AU$ and
for $r>58~AU$; the toroidal mass distribution can be detected at the
interval $15~AU < r < 52~AU$ and for $r > 56~AU$; finally, the
non-uniform disk can only be measured at $26~AU < r < 55~AU$. Notice
that, with the exception of the non-uniform disk mass distribution,
all remaining models can be essentially detected from distances beyond
$15 ~AU$.

\section{Conclusions}

Recent years have witnessed an increasing interest concerning the
Pioneer anomaly, elected by some as one of the withstanding enigmas of
contemporary physics. This increased exposure has prompted for various
proposals, of varying pertinence and scope. Asides from valid,
although mainly phenomenological theories, some have hailed that the
measured constant acceleration may be a manifestation of Modified
Newtonian Dynamics (MOND) \cite{MOND} in the Newtonian regime, or a
signal of perturbative corrections to the geodetic motion of test
particles, due to the presence of a cosmological constant; although
this was not the purpose of this study, the authors cannot refrain
from stating that, in our view, such claims stem solely from
fortuitous numerical coincidences.

Indeed, in the former case, the observed anomalous acceleration is of
the same order of magnitude (within a factor of eight) as the
fundamental acceleration cutoff of MOND, $a_P \sim a_0 \approx
10^{-10}~m/s^2$; however, this similitude is not supported by any
available study of the Newtonian regime of neither the
phenomenological MOND model, nor Tensor-Scalar-Vector theory (TeVeS),
its backing fundamental framework. Such an unsubstantiated claim
appears recursively in specialized and general discussions and media,
but the lack of proof or argument serves only to discredit it as a
viable modification or extension of General Relativity. Also relevant,
this suggestion is raised at a particularly troublesome time for MOND
and TeVeS alike, when the existence of Dark Matter is being directly
probed in galaxy clusters \cite{bullet}, and the viability of these
proposals is at stake, if not yet doomed \cite{COSPAR,Chiba}.

In the latter case, the identification of the Pioneer anomalous
acceleration as an effect induced by the expansion of the Universe is
doubly mischievous; firstly, most approaches rely on the apparent
relation $a_P \sim c H$, and attempt to derive this expression from
first principles, often with misguided arguments; indeed, the correct
use of the Schwarzschild-de Sitter metric, which extends the standard,
static and spherically symmetric metric, matching it with a de Sitter
boundary, yields an acceleration which is lower than the reported
value by over eleven orders of magnitude! Furthermore, and perhaps
more concerning, most of these claims fail to account for the simple
problem with this identification: an acceleration due to the expansion
of the Universe is, by nature, repulsive, while the anomalous
acceleration of the Pioneer probes is sunbound.

This said, one should exert this criticism with caution; indeed, while
the above proposals do not seem at all adequate, an immediate
dismissal of other, more evolved models could inhibit the development
of a phenomenologically viable extension of General Relativity, and
perhaps hinder a better understanding of gravitation
\cite{Bertolami}. Indeed, a confirmation of the existence of an
anomalous, constant acceleration in the Solar System could signal the
beginning of a new cycle in the pursuit for a clearer grasp of the
surrounding Universe. But, for now, one must proceed with care, and
further examine all possible ``natural'' causes for the phenomena,
with the before-mentioned modeling of thermal effects as a prime
candidate for inspection. Even if this comes out as the definitive
explanation for the anomaly, a highly valuable service will be done
for the scientific community, allowing for better deep space tracking,
more accurate ephemerides computation, and improved spacecraft design.

In either case, one thing is clear: prospects of new physics, any
mismodelling or engineering cause, the road towards a better
understanding of the Pioneer anomaly lies still ahead.


\begin{acknowledgments}

\samepage \noindent This work is partially supported by the Programa
Dinamizador de Ci\^{e}ncia e Tecnologia do Espa\c{c}o of the FCT -
Funda\c{c}\~{a}o para a Ci\^{e}ncia e Tecnologia (Portuguese Agency),
under the project PDCTE/FNU/50415/2003, and partially written while
attending the second Pioneer Anomaly Group Meeting at the
International Space Science Institute (ISSI) at Bern, from 19 to 23 of
February 2007. The authors would like to thank ISSI and its staff, for
hosting the group's meeting and accommodating for logistic
requirements. The work of JP is sponsored by the FCT under the grant
BPD 23287/2005.

\end{acknowledgments}


\end{document}